\documentclass[showpacs,amsmath,amssymb,twocolumn]{revtex4} 
\usepackage{mathrsfs}
\usepackage{epsfig}

\begin{document}

\title{Bleaching and diffusion dynamics in optofluidic dye lasers}

\author{Morten Gersborg-Hansen, S\o ren Balslev,\footnote{Present address: Ignis Photonyx
A/S, Blokken 84, DK-3460 Birker\o d, Denmark.} Niels Asger
Mortensen, and Anders Kristensen\footnote{Corresponding author:
ak@mic.dtu.dk; URL: www.optofluidics.dk, www.mic.dtu.dk/nil}}

\affiliation{MIC -- Department of Micro and Nanotechnology,
NanoDTU, Technical University of Denmark, Building 345east, \O
rsteds Plads, DK-2800 Kongens Lyngby, Denmark}

\date{\today}
\begin{abstract}
We have investigated the bleaching dynamics that occur in
optofluidic dye lasers where the liquid laser dye in a
microfluidic channel is locally bleached due to optical pumping.
We find that for microfluidic devices, the dye bleaching may be
compensated through diffusion of dye molecules alone. By relying
on diffusion rather than convection to generate the necessary dye
replenishment, our observation potentially allows for a
significant simplification of optofluidic dye laser device
layouts, omitting the need for cumbersome and costly external
fluidic handling or on-chip microfluidic pumping devices.
\end{abstract}


\maketitle

Optical techniques have proven powerful in chemical and
bio-chemical analysis. This has stimulated a large effort in
integrating fluidics and optics in lab-on-a-chip
microsystems~\cite{Verpoorte:2003}, partly defining the emerging
field of optofluidics, as recently reviewed by Psaltis \emph{et
al.}~\cite{Psaltis:2006} and Monat \emph{et
al.}~\cite{Monat:2007}. Among the investigated components are
miniaturized fluidic dye lasers, also referred to as optofluidic
dye
lasers~\cite{Helbo:2003,Balslev:2005a,GersborgHansen:2005,Vezenov:2005,Li:2006,GersborgHansen:2006}.
The so far reported optofluidic dye lasers are pulsed in order to
have a short interaction time between the dye molecules and the
pump light, thus suppressing the formation of triplet states
unsuitable for lasing. This contrasts macroscopic continuous-wave
dye lasers where the suppression is mediated by a jet flow of the
dye solution with typical velocities of several
m/s~\cite{Schaefer:1990}.

Dye bleaching resulting in limited lifetime is in general
considered a major disadvantage of organic dyes as active laser
medium. Typically, the problem of dye bleaching is addressed by
employing a continuous convective flow of liquid-dissolved dye
molecules, compensating the bleaching dynamics caused by the
external optical pump. The required convective dye-replenishing
flow has been achieved by external fluid-handling apparatus
(syringe
pumps)~\cite{Helbo:2003,Balslev:2005a,GersborgHansen:2005,Vezenov:2005,Li:2006},
an on-chip microfluidic pump~\cite{Galas:2005}, or by means of
capillary effect~\cite{GersborgHansen:2006,GersborgHansen:2007}.

\begin{figure}[b!]
\begin{center}
\epsfig{file=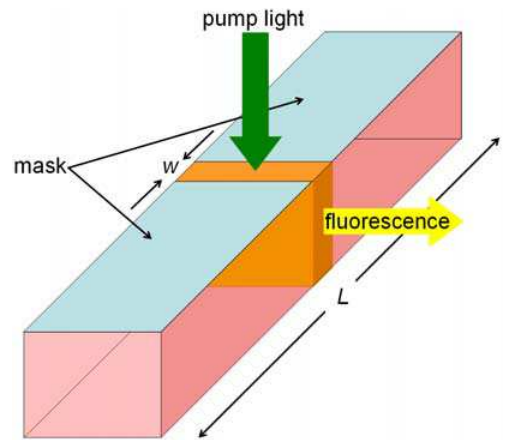, width=0.8\columnwidth,clip}
\end{center}
\caption{Experimental setup. Closed channel containing a liquid
solution of dye molecules. The dye molecules are optically pumped
by a pulsed, frequency-doubled Nd:YAG laser through a slit of
width $w$ covering the sample.}
\end{figure}

In this Letter we demonstrate that such optofluidic devices may
potentially be operated for days by diffusion without the need for
a convective flow. The key concept in this is very similar to the
entire paradigm behind miniaturized chemical-analysis systems
where scaling arguments are used to show the attractiveness of
micron-scale analytical devices compared to their macroscopic
counterparts. A recent review is given by Janasek \emph{et al.} in
Ref.~\cite{Janasek:2006}.

We have investigated the local dye bleaching dynamics which is
characteristic in optofluidic dye lasers. Based on our findings we
propose a dye replenishment mechanism which takes advantage of the
classical diffusion present when dye molecules are dissolved in a
liquid and placed in a microfluidic device. The bleaching of dye
molecules by the optical pump will introduce gradients in the
concentration of non-bleached dye molecules $c({\bf r},t)$, thus
activating the diffusion mechanism associated with the thermally
driven Brownian motion of the dye molecules. The
diffusion-bleaching dynamics is governed by a classical
diffusion-convection equation with an additional drain (sink) term

\begin{equation}\label{eq:diffusion}
 D \nabla^2 c({\bf r},t) =
\frac{\partial}{\partial t}\:c({\bf r},t)+{\bf v}\cdot
{\bf\nabla}c({\bf r},t)+\Gamma({\bf r})\:c({\bf r},t)
\end{equation}
where $\Gamma$ is the bleaching rate, $D$ is the diffusion
constant, and $\bf v$ is the velocity field of a possible flow.
For rhodamine 6G dye molecules dissolved in ethylene glycol, $D$
is estimated to $D\sim 1.5\times 10^{-11}\,{\rm m^2/s}$ by taking
an experimental value for rhodamine 6G molecules in
water~\cite{Rigler:1993} and scaling with the viscosity at
25$^\circ$C, using the Stokes-Einstein relation. On average, the
random walk transports a dye molecule approximately 0.1~mm in ten
minutes. For typical optical pumping levels and repetition rates
we estimate this to be sufficient to replenish bleached dye in a
miniaturized dye laser. This statement is supported by both
Eq.~(\ref{eq:diffusion}) as well as by our experimental studies of
a particular optofluidic device.

\begin{figure}[t!]
\begin{center}
\epsfig{file=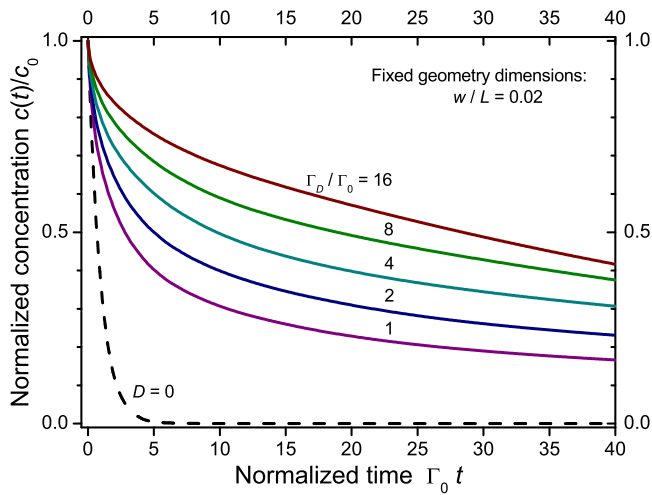, width=1\columnwidth,clip}
\end{center}
\caption{Plot of dye molecule concentration versus time in a
closed quasi one-dimensional microfluidic channel of length $L$,
see Fig.~1. Bleaching by the optical pump occurs in a narrow slit
of width $w=0.02\,L$ with a bleaching rate $\Gamma_0$. Diffusion
changes the dynamics away from the pure exponential bleaching
($D=0$) and yields a much slower decay.}
\end{figure}

To analyze the bleaching-diffusion dynamics, we use a simple,
idealized one-dimensional model system, resembling the
experimental setup outlined in Fig.~1. We consider a situation
where the liquid-dissolved dye molecules are optically pumped
through a narrow slit of width $w$ covering a microfluidic channel
of length $L$ connecting two ideal liquid reservoirs supporting an
always constant concentration $c_0$ (not shown on the figure).
With $x$ being along the direction of the quasi one-dimensional
channel we can thus make the approximation $\Gamma(x)\simeq
\Gamma_0 \, \Theta(w/2-|x|)$ where $\Theta(x)$ is the Heaviside
step function with $\Theta(x)=0$ for $x<0$ and $\Theta(x)=1$ for
$x>0$. In the symmetric case the concentration will of course
always have a minimum in the center of the slit so in the
following we focus on $x=0$ and suppress the spatial variable for
the sake of clarity.

The dynamics may conveniently be understood by dimensional
analysis. Since the characteristic timescale is set by the inverse
bleaching rate $\Gamma_0^{-1}$ and the characteristic length scale
by the slit width $w$, we readily arrive at a diffusion rate
$\Gamma_D= D/w^2$ and a convection rate $\Gamma_v=v/w$ which
should be compared to the externally controlled bleaching rate
$\Gamma_0$. In macroscopic dye lasers one has $\Gamma_v\gg
\Gamma_0\gg \Gamma_D$ so that un-bleached dye molecules are
supplied on a faster time scale than the bleaching. This strategy
has been central to the so far reported microfluidic dye
lasers~\cite{Helbo:2003,Balslev:2005a,GersborgHansen:2005,Vezenov:2005,Li:2006,Galas:2005,GersborgHansen:2006,GersborgHansen:2007}.
However, the very different scaling of $\Gamma_D$ and $\Gamma_v$
with $w$ offers an alternative and attractive replenishment
mechanism in micron-scale systems. Usually the convective term
${\mathbf v}\cdot {\mathbf \nabla} c$ is driving the
replenishment, but if $\Gamma_D\gg \max(\Gamma_v,\Gamma_0)$ we
have the freedom to completely turn off convection and entirely
rely on diffusion. In this work we will emphasize diffusion and
the convective term will be explicitly zero, i.e. ${\bf v}=0$.

The initial dynamics is characterized by the absence of gradients
and for $t\ll \Gamma_D^{-1}$, we get the expected solution
$c(t)\simeq c_0 \exp\left[-\Gamma_0 t\right]$. This exponential
dynamics, with a characteristic time scale $\Gamma_0^{-1}$, is the
major source of concern for the bleaching-defined lifetime of the
device outlined in Fig.~1. However, the bleaching, expressed as
$c\rightarrow 0$ for $t\gg \Gamma_0^{-1}$, may be compensated by
diffusion which becomes efficient for $t\gtrsim \Gamma_D^{-1}$.
The addition of rates (like Mathiesen's rule for addition of
scattering rates) suggests that for $t\gtrsim \Gamma_D^{-1}$ the
decay will be governed by $\Gamma_D$ rather than $\Gamma_0$.
Eq.~(\ref{eq:diffusion}) is difficult to solve analytically, but
the dynamics may easily be studied numerically in more detail.
Figure~2 illustrates the bleaching-diffusion dynamics for a
channel much longer than the slit width, $L\gg w$. As seen, the
above analysis accounts well for the qualitative behavior found
from time-dependent finite-element simulations of
Eq.~(\ref{eq:diffusion}) for varying values of $\Gamma_D$ relative
to $\Gamma_0$. Clearly, diffusion changes the dynamics away from
the pure exponential bleaching (lower curve) to a situation with a
much slower decay, thus potentially increasing the
bleaching-limited lifetime of the device dramatically.

Earlier experiments on dye-doped polymer lasers suggest that the
bleaching rate is of the order $\Gamma_0\sim 10^{-4}\,{\rm
s}^{-1}$ for typical concentrations and pumping
levels~\cite{Balslev:2005}. The diffusion rate $\Gamma_D$ would be
comparable to $\Gamma_0$ for a slit of macroscopic width $w\sim
1\,{\rm mm}$. For a micron-scale slit of width $w\sim 100\,{\rm
\mu m}$, however, the diffusion rate increases by one order of
magnitude to $\Gamma_D\sim 10^{-3}\,{\rm s}^{-1}$. These
order-of-magnitude estimates illustrate the importance of the
$w^{-2}$-scaling of $\Gamma_D$ and support our proposal of
miniaturized optofluidic dye lasers with $\Gamma_D\gg\Gamma_0$,
thus fully avoiding any need for external fluidic handling systems
such as syringe pumps or complicated on-chip pumping schemes.

The implicit hypothesis behind the above analysis is of course
that the dye laser output signal somehow correlates with the
concentration of un-bleached dye molecules. In order to verify our
predictions experimentally we use the intensity of the
fluorescence as an indirect measure of the local dye concentration
in the optically pumped volume. In the experiments, we filled a
flexible polymer tube of diameter $\simeq$~800~$\rm \mu m$ with a
$2\times10^{-4}~\rm{mol/L}$ solution of rhodamine 6G in ethylene
glycol. For this dye concentration level, the quantum efficiency
has a weak dependency on the dye
concentration~\cite{Arbeloa:1988}. The liquid volume was confined
to a length of $L = 8$~mm and the dye was locally pumped through a
slit of varying width $w$ using a pulsed, frequency doubled Nd:YAG
laser: wavelength 532~nm, pulse duration 5~ns, repetition rate
10~Hz, and average pulse energy fluence 32~$\rm \mu$J$/$mm$^2$.
The chosen pumping level is typical for optofluidic dye lasers
operating well above
threshold~\cite{Li:2006,GersborgHansen:2006,GersborgHansen:2007}.
The fluorescence signal indicated in Fig.~1 was collected at an
angle normal to the incident pump light by an optical fiber and
measured by a fixed grating spectrometer (resolution 0.15 nm).

\begin{figure}[t!]
\begin{center}
\epsfig{file=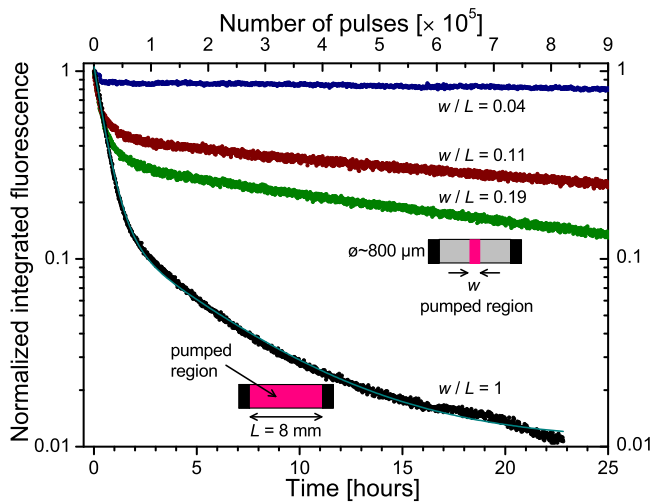, width=1\columnwidth,clip}
\end{center}
\caption{Temporal decay of fluorescence for different optical
pumping configurations. The lower curve ($w/L=1$) is for spatially
homogeneous pumping while the upper curves correspond to spatially
inhomogeneous pumping through slits of varying width $w$. In the
case of $w/L=0.04$, the fluorescence signal is nearly constant on
the timescale of days after a short initial decay (first 25 hours
shown).}
\end{figure}

Figure~3 shows the temporal decay of the integrated spectrally
broad fluorescence signal for different optical pumping
configurations. For spatially homogeneous optical pumping (no
slit, $w/L=1$), the data show an initial fast decay, followed by a
much slower decay with a characteristic decay time of 5.0~hours.
The behavior of an initial fast decay followed by a slow decay
does not arise from pump laser fluctuations or from changes in the
spectral output pattern of the device. When pumping through a slit
($w/L<1$), the fast decay remains and the rate of the slow decay
decreases with $w$. In the case of $w/L=0.04$ ($w\simeq 300\,{\rm
\mu m}$), a nearly stable output level on the timescale of days is
observed. The experimentally observed fluorescence dynamics is in
qualitative agreement with the bleaching-diffusion dynamics found
in our quasi one-dimensional model, see Fig.~2. We would like to
emphasize that such optofluidic devices provide a typical level of
output power that is more than sufficient for use as an on-chip
light source in lab-on-a-chip applications~\cite{Balslev:2006a}.

In conclusion, we have investigated the bleaching and diffusion
dynamics in optofluidic dye lasers caused by local bleaching of
the laser dye. A simple one-dimensional diffusion model was used
to explore the characteristic evolution of the local un-bleached
dye concentration in the optically pumped or bleached volume of
the device. In the absence of convective flow, the asymptotic
decay of the local dye concentration in the optically pumped
volume is governed by the diffusion rate and the lifetime is
mainly limited by the capacity of the fluidic reservoirs. Generic
microfluidic platforms typically allow for device layouts with a
large volume ratio between the fluidic reservoir and the region
being optically pumped. In order to put our proposal in
perspective, a reservoir with a volume of the order 1~cm$^3$, with
an optically pumped volume of $2.5\times10^{-7}~\rm{cm}^3$ and a
bleaching rate of $\Gamma_0\sim 10^{-4}\,{\rm
s}^{-1}$~\cite{Balslev:2005}, would in principle allow for
continuous operation for more than a thousand years for typical
pumping levels and repetition rates. These conclusions drawn from
the simple model are supported by basic experiments. Our
investigations reveal the possibility that such optofluidic dye
laser devices may potentially be operated for days by diffusion
without the need for a convective flow. Relying on diffusion
rather than convection to generate the necessary dye replenishment
significantly simplifies optofluidic dye laser device layouts,
omitting the need for cumbersome and costly external fluidic
handling or on-chip microfluidic pumping devices.

This work was supported by the \emph{Danish Technical Research
Council} (STVF, grant no: 26-02-0064) and the \emph{Danish Council
for Strategic Research} through the \emph{Strategic Program for
Young Researchers} (grant no: 2117-05-0037).


\begin{thebibliography}{16}
\expandafter\ifx\csname
natexlab\endcsname\relax\def\natexlab#1{#1}\fi
\expandafter\ifx\csname bibnamefont\endcsname\relax
  \def\bibnamefont#1{#1}\fi
\expandafter\ifx\csname bibfnamefont\endcsname\relax
  \def\bibfnamefont#1{#1}\fi
\expandafter\ifx\csname citenamefont\endcsname\relax
  \def\citenamefont#1{#1}\fi
\expandafter\ifx\csname url\endcsname\relax
  \def\url#1{\texttt{#1}}\fi
\expandafter\ifx\csname
urlprefix\endcsname\relax\def\urlprefix{URL }\fi
\providecommand{\bibinfo}[2]{#2}
\providecommand{\eprint}[2][]{\url{#2}}

\bibitem[{\citenamefont{Verpoorte}(2003)}]{Verpoorte:2003}
\bibinfo{author}{\bibfnamefont{E.}~\bibnamefont{Verpoorte}},
  \bibinfo{journal}{Lab Chip} \textbf{\bibinfo{volume}{3}}, \bibinfo{pages}{42N
  } (\bibinfo{year}{2003}).

\bibitem[{\citenamefont{Psaltis et~al.}(2006)\citenamefont{Psaltis, Quake, and
  Yang}}]{Psaltis:2006}
\bibinfo{author}{\bibfnamefont{D.}~\bibnamefont{Psaltis}},
  \bibinfo{author}{\bibfnamefont{S.~R.} \bibnamefont{Quake}}, \bibnamefont{and}
  \bibinfo{author}{\bibfnamefont{C.~H.} \bibnamefont{Yang}},
  \bibinfo{journal}{Nature} \textbf{\bibinfo{volume}{442}}, \bibinfo{pages}{381
  } (\bibinfo{year}{2006}).

\bibitem[{\citenamefont{Monat et~al.}(2007)\citenamefont{Monat, Domachuk, and
  Eggleton}}]{Monat:2007}
\bibinfo{author}{\bibfnamefont{C.}~\bibnamefont{Monat}},
  \bibinfo{author}{\bibfnamefont{P.}~\bibnamefont{Domachuk}}, \bibnamefont{and}
  \bibinfo{author}{\bibfnamefont{B.~J.} \bibnamefont{Eggleton}},
  \bibinfo{journal}{Nature Photonics} \textbf{\bibinfo{volume}{1}},
  \bibinfo{pages}{106 } (\bibinfo{year}{2007}).

\bibitem[{\citenamefont{Helbo et~al.}(2003)\citenamefont{Helbo, Kristensen, and
  Menon}}]{Helbo:2003}
\bibinfo{author}{\bibfnamefont{B.}~\bibnamefont{Helbo}},
  \bibinfo{author}{\bibfnamefont{A.}~\bibnamefont{Kristensen}},
  \bibnamefont{and} \bibinfo{author}{\bibfnamefont{A.}~\bibnamefont{Menon}},
  \bibinfo{journal}{J. Micromech. Microeng.} \textbf{\bibinfo{volume}{13}},
  \bibinfo{pages}{307 } (\bibinfo{year}{2003}).

\bibitem[{\citenamefont{Balslev and Kristensen}(2005)}]{Balslev:2005a}
\bibinfo{author}{\bibfnamefont{S.}~\bibnamefont{Balslev}} \bibnamefont{and}
  \bibinfo{author}{\bibfnamefont{A.}~\bibnamefont{Kristensen}},
  \bibinfo{journal}{Opt. Express} \textbf{\bibinfo{volume}{13}},
  \bibinfo{pages}{344 } (\bibinfo{year}{2005}).

\bibitem[{\citenamefont{Gersborg-Hansen
  et~al.}(2005)\citenamefont{Gersborg-Hansen, Balslev, Mortensen, and
  Kristensen}}]{GersborgHansen:2005}
\bibinfo{author}{\bibfnamefont{M.}~\bibnamefont{Gersborg-Hansen}},
  \bibinfo{author}{\bibfnamefont{S.}~\bibnamefont{Balslev}},
  \bibinfo{author}{\bibfnamefont{N.~A.} \bibnamefont{Mortensen}},
  \bibnamefont{and}
  \bibinfo{author}{\bibfnamefont{A.}~\bibnamefont{Kristensen}},
  \bibinfo{journal}{Microelectron. Eng.} \textbf{\bibinfo{volume}{78-79}},
  \bibinfo{pages}{185 } (\bibinfo{year}{2005}).

\bibitem[{\citenamefont{Vezenov et~al.}(2005)\citenamefont{Vezenov, Mayers,
  Conroy, Whitesides, Snee, Chan, Nocera, and Bawendi}}]{Vezenov:2005}
\bibinfo{author}{\bibfnamefont{D.~V.} \bibnamefont{Vezenov}},
  \bibinfo{author}{\bibfnamefont{B.~T.} \bibnamefont{Mayers}},
  \bibinfo{author}{\bibfnamefont{R.~S.} \bibnamefont{Conroy}},
  \bibinfo{author}{\bibfnamefont{G.~M.} \bibnamefont{Whitesides}},
  \bibinfo{author}{\bibfnamefont{P.~T.} \bibnamefont{Snee}},
  \bibinfo{author}{\bibfnamefont{Y.}~\bibnamefont{Chan}},
  \bibinfo{author}{\bibfnamefont{D.~G.} \bibnamefont{Nocera}},
  \bibnamefont{and} \bibinfo{author}{\bibfnamefont{M.~G.}
  \bibnamefont{Bawendi}}, \bibinfo{journal}{J. Am. Chem. Soc.}
  \textbf{\bibinfo{volume}{127}}, \bibinfo{pages}{8952} (\bibinfo{year}{2005}).

\bibitem[{\citenamefont{Li et~al.}(2006)\citenamefont{Li, Zhang, Emery,
  Scherer, and Psaltis}}]{Li:2006}
\bibinfo{author}{\bibfnamefont{Z.~Y.} \bibnamefont{Li}},
  \bibinfo{author}{\bibfnamefont{Z.~Y.} \bibnamefont{Zhang}},
  \bibinfo{author}{\bibfnamefont{T.}~\bibnamefont{Emery}},
  \bibinfo{author}{\bibfnamefont{A.}~\bibnamefont{Scherer}}, \bibnamefont{and}
  \bibinfo{author}{\bibfnamefont{D.}~\bibnamefont{Psaltis}},
  \bibinfo{journal}{Opt. Express} \textbf{\bibinfo{volume}{14}},
  \bibinfo{pages}{696 } (\bibinfo{year}{2006}).

\bibitem[{\citenamefont{Gersborg-Hansen and
  Kristensen}(2006)}]{GersborgHansen:2006}
\bibinfo{author}{\bibfnamefont{M.}~\bibnamefont{Gersborg-Hansen}}
  \bibnamefont{and}
  \bibinfo{author}{\bibfnamefont{A.}~\bibnamefont{Kristensen}},
  \bibinfo{journal}{Appl. Phys. Lett.} \textbf{\bibinfo{volume}{89}},
  \bibinfo{pages}{103518} (\bibinfo{year}{2006}).

\bibitem[{\citenamefont{Sch\"afer}(1990)}]{Schaefer:1990}
\bibinfo{author}{\bibfnamefont{F.~P.} \bibnamefont{Sch\"afer} (ed.)},
  \emph{\bibinfo{title}{Dye Lasers}} (\bibinfo{publisher}{Springer-Verlag},
  \bibinfo{address}{Berlin Heidelberg}, \bibinfo{year}{1990}),
  \bibinfo{edition}{3rd} ed.

\bibitem[{\citenamefont{Galas et~al.}(2005)\citenamefont{Galas, Torres,
  Belotti, Kou, and Chen}}]{Galas:2005}
\bibinfo{author}{\bibfnamefont{J.~C.} \bibnamefont{Galas}},
  \bibinfo{author}{\bibfnamefont{J.}~\bibnamefont{Torres}},
  \bibinfo{author}{\bibfnamefont{M.}~\bibnamefont{Belotti}},
  \bibinfo{author}{\bibfnamefont{Q.}~\bibnamefont{Kou}}, \bibnamefont{and}
  \bibinfo{author}{\bibfnamefont{Y.}~\bibnamefont{Chen}},
  \bibinfo{journal}{Appl. Phys. Lett.} \textbf{\bibinfo{volume}{86}},
  \bibinfo{pages}{264101} (\bibinfo{year}{2005}).

\bibitem[{\citenamefont{Gersborg-Hansen and
  Kristensen}(2007)}]{GersborgHansen:2007}
\bibinfo{author}{\bibfnamefont{M.}~\bibnamefont{Gersborg-Hansen}}
  \bibnamefont{and}
  \bibinfo{author}{\bibfnamefont{A.}~\bibnamefont{Kristensen}},
  \bibinfo{journal}{Opt. Express} \textbf{\bibinfo{volume}{15}},
  \bibinfo{pages}{137 } (\bibinfo{year}{2007}).

\bibitem[{\citenamefont{Janasek et~al.}(2006)\citenamefont{Janasek, Franzke,
  and Manz}}]{Janasek:2006}
\bibinfo{author}{\bibfnamefont{D.}~\bibnamefont{Janasek}},
  \bibinfo{author}{\bibfnamefont{J.}~\bibnamefont{Franzke}}, \bibnamefont{and}
  \bibinfo{author}{\bibfnamefont{A.}~\bibnamefont{Manz}},
  \bibinfo{journal}{Nature} \textbf{\bibinfo{volume}{442}}, \bibinfo{pages}{374
  } (\bibinfo{year}{2006}).

\bibitem[{\citenamefont{Rigler et~al.}(1993)\citenamefont{Rigler, Mets,
  Widengren, and Kask}}]{Rigler:1993}
\bibinfo{author}{\bibfnamefont{R.}~\bibnamefont{Rigler}},
  \bibinfo{author}{\bibfnamefont{{\"U}.}~\bibnamefont{Mets}},
  \bibinfo{author}{\bibfnamefont{J.}~\bibnamefont{Widengren}},
  \bibnamefont{and} \bibinfo{author}{\bibfnamefont{P.}~\bibnamefont{Kask}},
  \bibinfo{journal}{Eur. Biophys. J. Biophys. Lett.}
  \textbf{\bibinfo{volume}{22}}, \bibinfo{pages}{169 } (\bibinfo{year}{1993}).

\bibitem[{\citenamefont{Balslev et~al.}(2005)\citenamefont{Balslev, Rasmussen,
  Shi, and Kristensen}}]{Balslev:2005}
\bibinfo{author}{\bibfnamefont{S.}~\bibnamefont{Balslev}},
  \bibinfo{author}{\bibfnamefont{T.}~\bibnamefont{Rasmussen}},
  \bibinfo{author}{\bibfnamefont{P.}~\bibnamefont{Shi}}, \bibnamefont{and}
  \bibinfo{author}{\bibfnamefont{A.}~\bibnamefont{Kristensen}},
  \bibinfo{journal}{J. Micromech. Microeng.} \textbf{\bibinfo{volume}{15}},
  \bibinfo{pages}{2456 } (\bibinfo{year}{2005}).

\bibitem[{\citenamefont{Arbeloa et~al.}(1988)\citenamefont{Arbeloa, Ojeda, and
  Arbeloa}}]{Arbeloa:1988}
\bibinfo{author}{\bibfnamefont{F.~L.} \bibnamefont{Arbeloa}},
  \bibinfo{author}{\bibfnamefont{P.~R.} \bibnamefont{Ojeda}}, \bibnamefont{and}
  \bibinfo{author}{\bibfnamefont{I.~L.} \bibnamefont{Arbeloa}},
  \bibinfo{journal}{J. Photochem. Photobiol. A: Chem.}
  \textbf{\bibinfo{volume}{45}}, \bibinfo{pages}{313} (\bibinfo{year}{1988}).

\bibitem[{\citenamefont{Balslev et~al.}(2006)\citenamefont{Balslev, Jorgensen,
  Bilenberg, Mogensen, Snakenborg, Geschke, Kutter, and
  Kristensen}}]{Balslev:2006a}
\bibinfo{author}{\bibfnamefont{S.}~\bibnamefont{Balslev}},
  \bibinfo{author}{\bibfnamefont{A.~M.} \bibnamefont{Jorgensen}},
  \bibinfo{author}{\bibfnamefont{B.}~\bibnamefont{Bilenberg}},
  \bibinfo{author}{\bibfnamefont{K.~B.} \bibnamefont{Mogensen}},
  \bibinfo{author}{\bibfnamefont{D.}~\bibnamefont{Snakenborg}},
  \bibinfo{author}{\bibfnamefont{O.}~\bibnamefont{Geschke}},
  \bibinfo{author}{\bibfnamefont{J.~P.} \bibnamefont{Kutter}},
  \bibnamefont{and}
  \bibinfo{author}{\bibfnamefont{A.}~\bibnamefont{Kristensen}},
  \bibinfo{journal}{Lab Chip} \textbf{\bibinfo{volume}{6}}, \bibinfo{pages}{213
  } (\bibinfo{year}{2006}).

\end{thebibliography}

\end{document}